\documentclass[preprint,12pt]{aastex}  %--original for ApJ
\usepackage{epsfig}

\def\fei{Fe\,{\sc i}}

\def\fexvii{Fe\,{\sc xvii}}

\def\fexxv{Fe\,{\sc xxv}}
\def\fexxvi{Fe\,{\sc xxvi}}

\def\civ{C\,{\sc iv}}

\def\mathv{\textbf{\em v}}
\def\mathB{\textbf{\em B}}

\def\cm{\ifmmode {\rm cm}^{-1} \else cm$^{-1}$ \fi}
\def\s{\ifmmode {\rm s}^{-1} \else s$^{-1}$ \fi}
\def\cc{\ifmmode {\rm cm}^{-3} \else cm$^{-3}$ \fi}
\def\cs{\ifmmode {\rm cm}^{-2} \else cm$^{-2}$ \fi}
\def\g{\ifmmode \gamma \else $\gamma$\fi}
\def\G{\ifmmode \Gamma \else $\Gamma$\fi}
\def\Gs{\ifmmode \Gamma~ \else $\Gamma~$\fi}

\def\gc{\ifmmode \gamma_{\rm c} \else $\gamma_{\rm c}$ \fi}
\def\sw{Schwarzschild~}
\def\gsim{\mathrel{\raise.5ex\hbox{$>$}\mkern-14mu
             \lower0.6ex\hbox{$\sim$}}}
\def\lsim{\mathrel{\raise.3ex\hbox{$<$}\mkern-14mu
             \lower0.6ex\hbox{$\sim$}}}
\def\simless{\mathbin{\lower 3pt\hbox
     {$\rlap{\raise 5pt\hbox{$\char'074$}}\mathchar"7218$}}}   %< or of order
\def\simmore{\mathbin{\lower 3pt\hbox
     {$\rlap{\raise 5pt\hbox{$\char'076$}}\mathchar"7218$}}}   %> or of order
\def\Msun{M_\odot}                                % solar masses
\def\4u{4U 1728--34}
\def\deg{^\circ}
\def\aa{\buildrel _{\circ} \over {\mathrm{A}}}

\lefthead{et al.} \righthead{\4u}

\shorttitle{High-Velocity Outflows in QSOs}

\shortauthors{Fukumura et al.}

\begin{document}

\title{Modeling High-Velocity QSO Absorbers with Photoionized MHD Disk-Winds }

%\date{\today}

%
\author{\textsc{Keigo Fukumura}\altaffilmark{1,2,3},
\textsc{Demosthenes Kazanas}\altaffilmark{3},
\textsc{Ioannis Contopoulos}\altaffilmark{4}, \\
\textsc{and} \\
\textsc{Ehud Behar}\altaffilmark{5} }

\altaffiltext{1}{Email: Keigo.Fukumura@nasa.gov}
\altaffiltext{2}{University of Maryland, Baltimore County
(UMBC/CRESST), Baltimore, MD 21250} \altaffiltext{3}{Astrophysics
Science Division, NASA/Goddard Space Flight Center, Greenbelt, MD
20771} \altaffiltext{4}{Research Center for Astronomy, Academy of
Athens, Athens 11527, Greece} \altaffiltext{5}{Department of
Physics, Technion, Haifa 32000, Israel}

\begin{abstract}

\baselineskip=15pt

We extend our modeling of the ionization structure of
magnetohydrodynamic (MHD) accretion-disk winds, previously applied
to Seyfert galaxies, to a population of quasi-stellar-objects (QSOs)
of much lower X-ray-to-UV flux ratios, i.e. smaller $\alpha_{\rm
ox}$ index, motivated by
%
%an increasing number of
%
UV/X-ray ionized
absorbers with extremely high outflow velocities in UV-luminous
QSOs. We demonstrate that magnetically-driven winds ionized by a
spectrum with $\alpha_{\rm ox} \simeq -2$ can produce the charge
states responsible for \civ ~and \fexxv/\fexxvi ~absorption in wind
regions with corresponding maximum velocities of $v$(\civ) $\lsim
0.1c$ and $v({\rm \fexxv}) \lsim 0.6 c$ (where $c$ is the speed of
light) and column densities $N_H \sim 10^{23}-10^{24}$ cm$^{-2}$, in
general agreement with observations. In contrast to the conventional
radiation-driven wind models, {\it high-velocity flows are always
present in our MHD-driven winds} but manifest in the absorption
spectra only for $\alpha_{\rm ox} \lsim -2$, as larger $\alpha_{\rm
ox}$ values ionize the wind completely out to radii too large to
demonstrate the presence of these high velocities. We thus predict
increasing velocities of these ionized absorbers with decreasing
(steeper) $\alpha_{\rm ox}$, a quantity that emerges as the defining
parameter in the kinematics of the AGN UV/X-ray absorbers.
%

%The physical reason for this effect is the scaling of velocity and ionization with position in the MHD wind, and has nothing to do with radiation pressure. Steep $\alpha_{\rm ox}$ allows the observed charge states to form closer to the center than in Seyferts, and where the MHD wind is faster.

%We propose these photoionized MHD winds as the fundamental
%ingredient of the UV and X-ray AGN absorption features.

\end{abstract}


\baselineskip=15pt

\section{Introduction}

%The launch of {\it ASCA}, {\it Chandra} and {\it XMM-Newton} ushered a new era in
%X-ray astronomy with the discovery of absorption features in the
%spectra of AGN code named warm absorbers.
%
The launch of {\it Chandra} and {\it XMM-Newton}
%
%and the grating spectrometers on board
%
ushered a new era in X-ray astronomy of AGN outflows with the
discovery of absorption lines in the spectra that enabled for the
first time accurate charge state and velocity measurements. The long
observations of a number of AGNs revealed transitions of charge
states as diverse as \fei ~through \fexxvi. Since any atomic gas
with bound electrons absorbs X-rays these ions span a range of $\sim
10^5$ in ionization parameter\footnote[1]{$\xi \equiv L/(n r^2)$
where $L$ is an ionizing luminosity (between 1 and 1000 Ryd), $n$ is
the plasma number density and $r$ is distance from the ionizing
source.} $\xi$, a fact that underscores the great utility of X-ray
spectroscopy.
%
%Since any atomic gas with bound electrons absorbs x-rays, within 1.5 decades in frequency
%it encompasses transitions that span a factor of $10^5$ in $\xi$.
% (Keigo, this is a model dependent statement better stated in the next paragraph)
% and allows one to probe the properties of the absorbing medium along the
% observer's line of sight (LoS) over many decades in radius.

In a subsequent development, \citet{HBK07} and \citet[][hereafter,
B09]{B09} developed a statistical measure of the plethora of the
transitions in the {\it Chandra/XMM} spectra, the absorption measure
distribution (AMD), namely the differential hydrogen-equivalent
column $N_H$ of specific ions per decade of $\xi$, i.e. AMD $\equiv
d N_{\rm H}/d \log \xi$. Moreover, the AMD was found to be roughly
constant, i.e. $N_{\rm H}$ to be roughly independent of $\xi$, in
the small number of Seyferts for which the data quality allowed a
quantitative analysis. The functional form of the AMD is significant
as it can provide the plasma density along the observer's line of
sight (LoS), which for constant AMD is $n(r) \propto r^{-1}$.

%Unlikely as this function might at first appear as a wind density
%profile, it is one amongst the profiles of the 2D MHD disk-wind
%solutions of \citet[][hereafter, CL94]{CL94}. Winds of this type and
%specific density profile have also been used by \cite{KK94} to
%account for the IR properties of AGN and were proposed as
%alternatives to the AGN ``torus" (the large scale toroidal structure
%thought responsible for the AGN unification on the basis of the
%observer's inclination angle) because of their much higher
%equatorial than polar direction columns.

Motivated by the AMD systematics, \citet[][hereafter, FKCB]{FKCB}
employed the photoionization code \verb"XSTAR" \citep[][]{KB01} to
determine the ionization structure of the 2D winds of
\citet[][hereafter, CL94]{CL94} which provide for density profiles
such as $n \propto r^{-1}$. This density dependence on $r$ yields
also $\xi \propto r^{-1}$, thereby allowing for ionic species of
decreasing ionization with distance, but of columns similar to those
of high ionization. Importantly, these models are scale free: with
the radial coordinate $r$ normalized to the \sw radius, $r_s$, and
the mass flux to the Eddington rate, $\xi$ is independent of the
black hole mass $M$, implying broad applicability in galactic and
extragalactic settings. Assuming an ionizing spectral energy density
(SED) of $F_{\nu} \propto \nu^{-1}$ between 1 and 1000 Ryd, these
models were successful in reproducing the observed: (i) Slow
velocities ($v \sim 100-300$ km~s$^{-1}$) for the low ionization
transitions like \fexvii ~and fast outflows ($v \sim 1,000-3,000$
km~s$^{-1}$) for the high ionization ones such as \fexxv, and (ii)
AMD almost independent of $\xi$ for $-1 \lsim \log \xi \lsim 4$, in
agreement with the results of B09.

While X-ray absorption lines in Seyfert spectra are rather recent
discoveries, UV absorption lines in Seyferts and QSOs have been
known \citep[e.g.,][]{Crenshaw03,Brandt00}.
% Keigo, 2000 and 2003 is not very long ago
Also known since the earlier {\it ROSAT} surveys
\citep[e.g.][]{Kopko94,GreenMathur96} is that the X-ray-to-UV flux
ratio of the broad absorption line (BAL) QSOs [i.e. QSOs with blue
absorption \civ ~and Ly$\alpha$ troughs of $\Delta v/c \sim
0.04-0.1$ \citep[e.g.][]{HewettFoltz03,SP00}] is smaller than that
of the QSO majority, possibly due to absorption of the X-rays by the
BAL plasma. Indeed, this was confirmed by the {\it ASCA} detection
of high X-ray absorption column $N_H \geq 5 \times 10^{23}$
cm$^{-2}$ \citep[][]{Gallagher99}.
%
%In an effort to determine better
%the X-ray properties of BAL QSOs,
%
\citet[][hereafter, G06]{Gallagher06} later conducted a {\it
Chandra} survey combined with known UV absorption properties that
supported the earlier claims.
%
%that the X-ray weakness of the BAL QSOs is
%most likely due to an intrinsic absorber of high column.
%
The {\it Chandra} data of BAL QSOs indicate that $\alpha_{\rm
ox}(\rm BAL) \simeq -2.21$ (G06) is smaller than the mean QSO value
$\alpha_{\rm ox}(\rm mean) \simeq -2.0$ \footnote[2]{The spectral
index $\alpha_{\rm ox} \equiv 0.384 \log (f_{\rm 2keV} / f_{\rm
2500})$ measures the X-ray-to-UV relative brightness where $f_{\rm
2keV}$ and $f_{\rm 2500}$ are respectively 2 keV and $2500 \aa$ flux
densities \citep{Tananbaum79}}.
%In favor of such an interpretation argues also the
This result is augmented by a correlation between $\alpha_{\rm ox}$
and the $1-5$ keV X-ray photon index $\Gamma$:
%While the typically observed X-ray spectrum has $\Gamma \simeq 2$,
increased photoelectric absorption of  soft X-rays (i.e. smaller
$\alpha_{\rm ox})$ also yields a smaller effective $\Gamma$, as
observed.

The high outflow velocities of the prominent UV resonance lines
(\civ ~and Ly$\alpha$) in BAL QSOs were traditionally ascribed to
radiation-driven winds \citep{Weymann91}, in analogy with the winds
of O stars \citep{cak} and were modeled as such
\citep[e.g.][hereafter, MCGV]{murray95}. MCGV recognized and
included heuristically the effects of wind ionization and its
shielding from the QSO X-rays, a crucial process as ionization
reduces severely the effectiveness of line driving.
\citet[][hereafter, PSK]{PSK00} presented 2D hydrodynamic
simulations of these winds, including X-ray ionization, showing that
the required shielding is provided by the section of the wind
closest to the X-ray source that ``failed" to launch by being too
highly ionized, thereby allowing exterior segments to achieve
velocities in agreement with \civ ~observations.
%
%(high-velocity UV absorbers in QSO J2123-0050, however, seems to contradict with this view; Hamann, in private communication {\bf Keigo, this is too terse a comment, I suggest either elaborate, or omit if space does not allow}).

However, recent X-ray observations of BAL QSOs revealed absorption
features in their spectra identified with highly ionized
\fexxv/\fexxvi ~of column density $N_H \sim 10^{23}-10^{24}$
cm$^{-2}$, blueshifted to high velocities $v/c \sim 0.4-0.7$ (e.g.
APM~08279+5255, PG~1115+080 and H~1413+117)
% for APM~08279+5255
%; \cite{Chartas03,Chartas07}
%for PG~1115+080
%; \citealt{Chartas07}),
% for H~1413+117)
indicating that X-ray ionization does not necessarily inhibit
outflows, which can occur at velocities even higher than those seen
in the UV lines (e.g., \citealt{Chartas02,Chartas03,Chartas07};
\citealt{Chartas09}, hereafter, C09). Additional X-ray studies have
revealed a number of non-BAL QSOs
%[i.e. 500 km~s$^{-1}$ $\lsim \Delta v ($\civ$) \lsim$ 2,000 km~s$^{-1}$]
%Keigo, some of these have no UV absorption, so better not to specify velocities
that also exhibit similar X-ray absorbers at high velocities $v/c \sim 0.1-0.5$ (e.g.,
\citealt{Pounds03} %for PG~0844+349
; \citealt{Reeves03}% for PDS~456
%; \citealt{PP06} %for PG~1211+143
%; \citealt{ZW08}),
%for {\it Chandra}Deep Field South QSOs)
%instead I added Reeves et al. 2009, which I think is the best evidence (Ehud)
; \citealt{Reeves09}), while in APM~08279+5255 C09 have also noted a
correlation between $\Gamma$ and the velocity of \fexxv.
%However, the limited sample size of this type of
%observation does not allow one to draw strong conclusions concerning
%the relation of high-velocity X-ray absorbers to the high-velocity
%UV BALs.
%

Motivated by these observations, we examine in this letter the
conditions under which the magnetically-driven winds discussed in
FKCB can reproduce the observed velocities of the BAL QSO X-ray
features (\fexxv) {\em along with} those of their more common UV
transitions (\civ). In \S 2 we summarize the physics of MHD
accretion disk winds and the differences of the ionization
properties between Seyferts and BAL QSOs. In \S 3 we present our
results and demonstrate a number of well-defined correlations among
their kinematics, column, spectral index, and LoS angle and we
conclude with a summary and discussion in \S 4.

\section{The MHD Disk-Wind Model}

In this section we present a brief outline of the MHD winds,
originally formulated by Blandford \& Payne (1982) and generalized
by CL94 to include arbitrary distribution of axial current with
radius. Here and in FKCB we focus on the current distribution that
produces a density profile $n(r) \propto r^{-1}$, crucial for
obtaining the observed AMD behavior. The same distribution leads
also to a toroidal field $B_{\phi} \propto r^{-1}$ that has
equal magnetic energy per decade of (cylindrical) radius.

%The separation of variables of the MHD equations is made by
Self-similarity is assumed, i.e. power-law radial dependence for
all variables and solution of the remaining angular part of these
equations.
%
%This is not a very restrictive condition considering the
%large number of decades spanned by the AGN accretion flows.
As discussed in FKCB, this assumption is not very restrictive and justified {\em a posteriori}
by the large number of decades of $\xi$ in the AMD form.

The fundamental quantity of axisymmetric MHD is the magnetic stream
function  $\Psi(r,\theta)$, assumed to have the form $\Psi(r,\theta)
\equiv (r/r_o)^q \tilde{\Psi}(\theta) \Psi_o$, with $\Psi_o$ the
poloidal magnetic flux through the fiducial wind launch radius at
$r=r_o$. $\tilde{\Psi}(\theta)$ is its angular dependence to be
solved for and $q \simeq 1$ a free parameter that determines the
radial dependence of the poloidal current.
The scalings of the poloidal magnetic stream function carry over to
the rest of the wind properties of which we show only the magnetic
field, velocity and density (see FKCB):
\begin{eqnarray}
\mathB(r,\theta) &\equiv& (r/r_o)^{q-2} \tilde{\mathB}(\theta)B_o \ ,
\label{eq:eos} \\
\mathv(r,\theta) &\equiv& (r/r_o)^{-1/2} \tilde{\mathv}(\theta)v_o \ ,
 \label{eq:eos} \\
%     p(r,\theta) &\equiv& (r/r_o)^{2q-4} {\cal P}(\theta)B_o^2 \ ,
% \label{eq:pres2}  \\
n(r,\theta) &\equiv& (r/r_o)^{2q-3} \tilde{n}(\theta)B_o^2
v_o^{-2} m_p^{-1}\ ,
\end{eqnarray}
%
%.
where $m_p$ is the proton mass. The dimensionless angular functions
denoted by {\it tilde} must be obtained from the conservation
equations and the solution of the Grad-Shafranov equation with
initial values on the disk (denoted by the subscript ``o") at
($r=r_o, \theta = 90\deg$).
%
%It is more instructive to express the density normalization
%$(B_o/v_o)^2 m_p^{-1}$ in terms of the accretion or wind outflow
%rate $\dot m$, normalized to its Eddington value, as discussed in
%FKCB; then
%
The density normalization at ($r_o,90\deg$), setting
$\tilde{n}(90^{\circ})=1$, is given in terms of dimensionless
mass-accretion rate $\dot{m}$ (see FKCB) by
\begin{equation}
n_o = \frac{\eta_W \dot m}{2\sigma_T r_s}\ ,
\label{eq:dnorm}
\end{equation}
where $\eta_W$ is the ratio of the mass-outflow rate in the wind to
$\dot{m}$, assumed here to be of order unity
%fraction of the total mass flow that goes
%to the wind, typically,
and $\sigma_T$ is the Thomson cross-section. It is important to note
that because the mass flux in these winds depends in general on the
radius, $\dot m$ always refers to the mass flux at the innermost flow
radius at $r \simeq r_s$ where $r_s$ is the \sw radius. In the
present treatment we adopt the value $q=0.93$ resulting in
$n \propto r^{-1.14}$, the steepest density dependence on $r$ implied by the AGN AMD data of B09,
and in order to allow for the somewhat higher observed X-ray column than UV column.

With the dimensionless, mass-invariant wind structure
%
%nearly identical to that of FKCB
%
(see Fig.~1a)
%
%and independent of the accreting object mass $M$
%
for given $\dot m$ and $\theta$, the only significant difference
in the wind ionization properties across objects of different
luminosity is the spectral distribution of ionizing radiation. While
in FKCB we used a spectrum of the form $F_{\nu} \propto \nu^{-1}$ \citep[e.g.][]{Sim08,Sim10}, more appropriate for Seyferts,
here we add a bright UV disk source.
%It is known that $\alpha_{\rm ox}$ becomes more negative with increasing
%UV luminosity \citep{steffen05}.
The spectrum used in the present work is shown in Figure~1b; it
comprises a multicolor-disk (MCD) with an innermost temperature of 5
eV and an X-ray power-law (PL) of photon index $\Gamma$ normalized
by $\alpha_{\rm ox}$ \citep[e.g.][]{Everett05,Sim05}. We do
not include a soft X-ray excess
%
%\footnote[3]{With this component added we have noted that the modeled \civ\ velocity is reduced while it has little effect on the X-ray outflow velocity.}
%
in the SED, a feature more appropriate for narrow-line Seyfert
spectra \citep[e.g.][]{Pounds03,Sim05}. The PL has a low energy
cut-off at 5 eV and a high energy one at 200 keV. The total (X-ray
plus UV) luminosity is $L = 3 \times 10^{45}$ erg~s$^{-1}$.
%
%{\bf
%(Keigo, I assume this is the total luminosity, not only the power law.
%You did not change the luminosity for different $\alpha_{\rm ox}$, right?) }
%
%with photon index $\Gamma$ obtaining values between 1.5 and 2.5.
%While some QSOs
%showing high-velocity outflows appear to exhibit a soft X-ray
%excess, possibly due to a hotter part of the accretion disk
%\citep[e.g.][]{Pounds03,Reeves03}, our current treatment ignores
%this component to keep the SED as simple as possible \citep[see,
%however,][]{Sim05}.
%
%This simplified spectrum is similar to that used in other models of AGN outflows
%\citep[e.g.][]{Everett05,Sim05,Sim08,Sim10}.

\begin{figure}[t]% ------------------------------------- Figure~1
\begin{center}$
\begin{array}{cc}
\includegraphics[trim=0in 0in 0in
0in,keepaspectratio=false,width=2.7in,angle=-0,clip=false]
{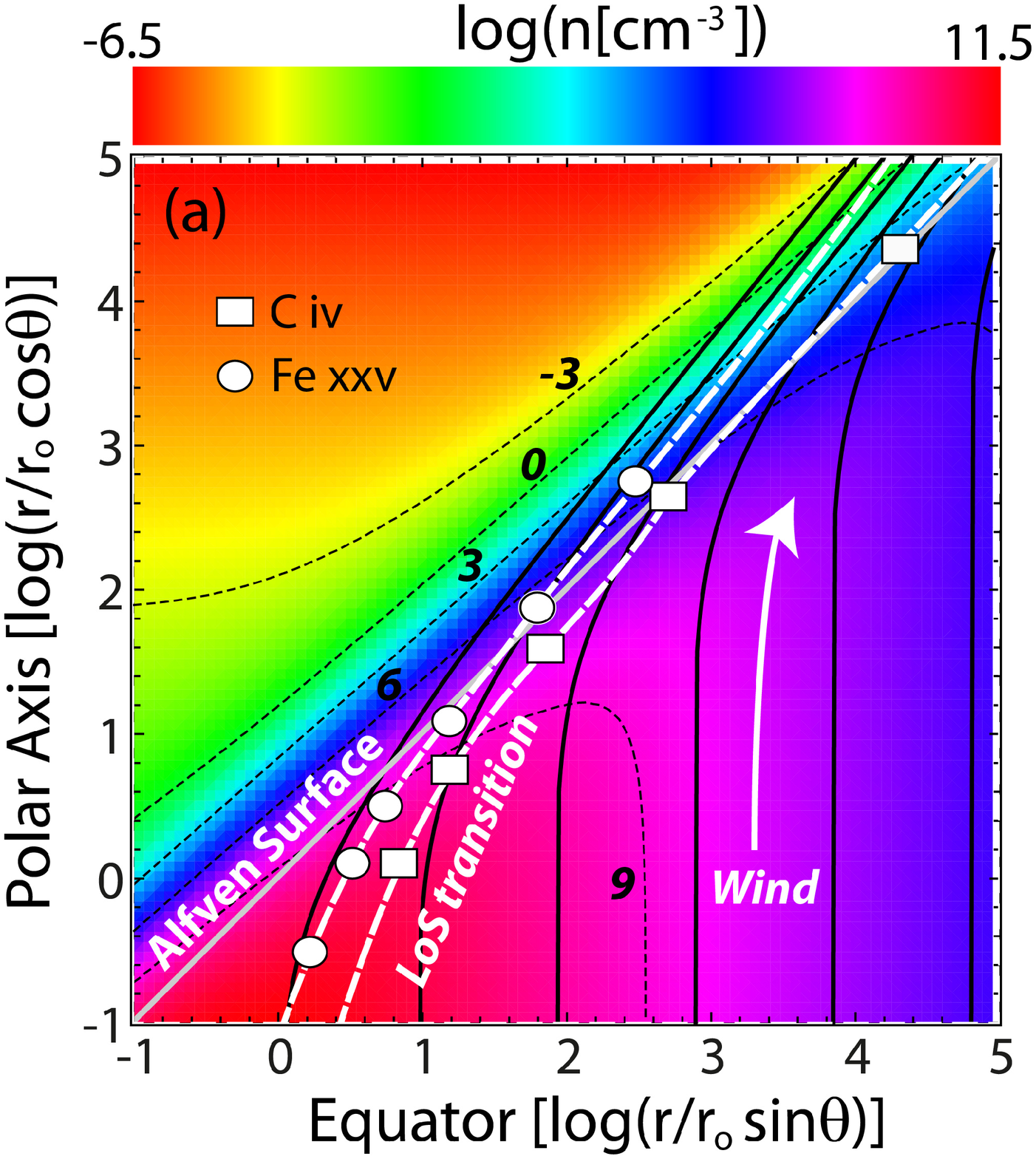} &
\includegraphics[trim=0in 0in 0in
0in,keepaspectratio=false,width=3.3in,angle=-0,clip=false]
{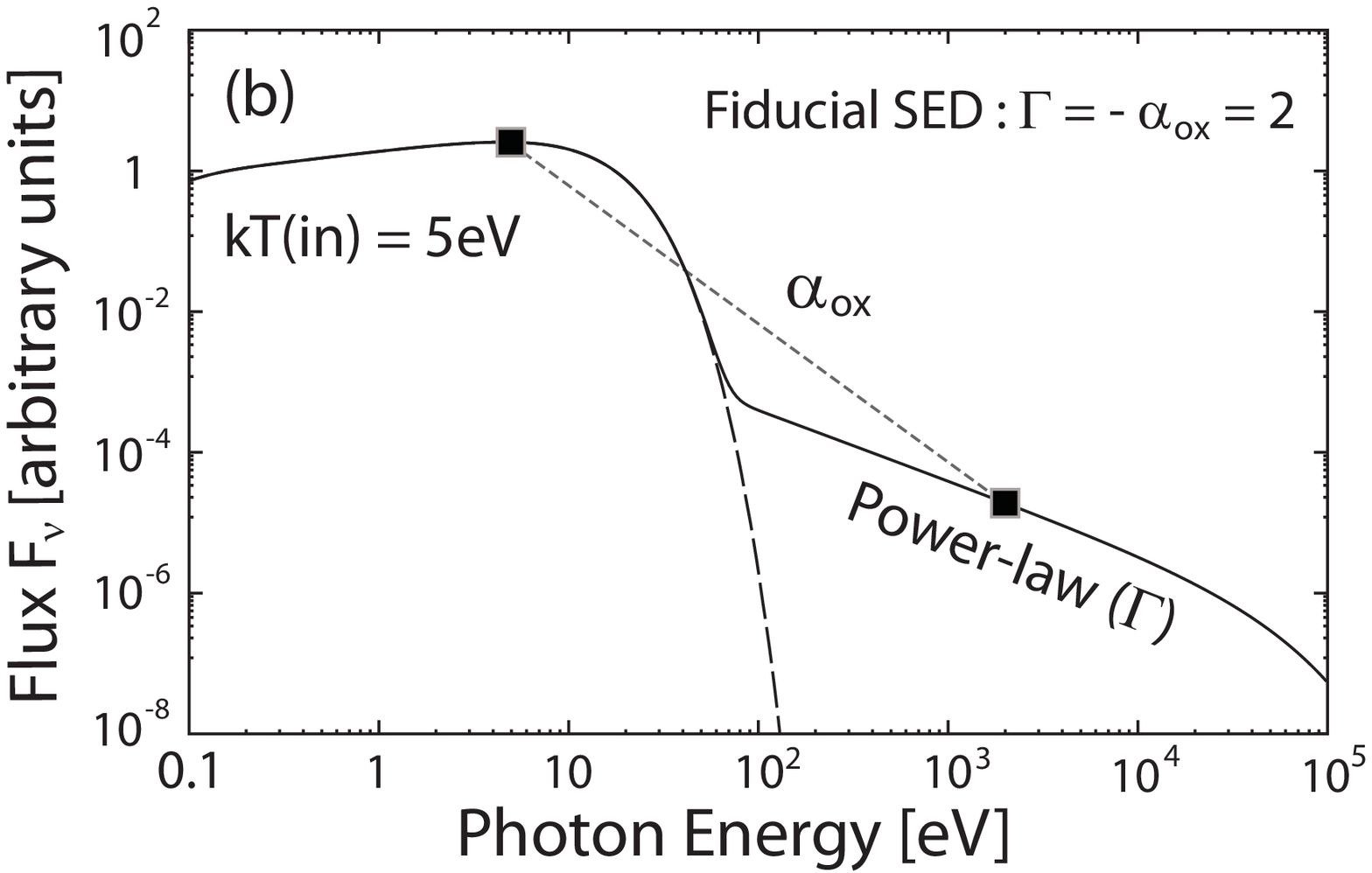}
\end{array}$
\end{center}
\caption{(a) Poloidal density structure $\log (n[\textrm{cm}^{-3}])$
of  MHD wind with $\dot{m}=0.5$; density contour curves ({\it
dotted} lines with numbers) and magnetic field
lines ({\it solid} curves) for $M=10^9 \Msun$. Also shown are the
positions of the \civ ~(${\rm square}=\square$) and \fexxv ~(${\rm circle}=\circ$)
which shift outward along LoS of decreasing $\theta$ [from $80\deg$ (innermost) to
$30\deg$ (outermost) by a $10\deg$ increment]. Note that the \civ~ position for
$\theta=30\deg$ lies outside the figure range.
%
%Square ($\square$) and circle ($\circ$) symbols respectively denote the position of \civ ~and \fexxv ~along the $\theta=80\deg, 70\deg, 60\deg, 50\deg, 40\deg$ and $30\deg$ LoS.
%
(b) The form of the assumed input SED consisting of a thermal MCD of
innermost temperature $k T_{\rm in}=5$ eV and a PL
continuum of photon index $\Gamma$ normalized by $\alpha_{\rm
ox}$. } \label{fig:fig1}
\end{figure}

\section{Results}

With the background flow ($\dot m = 0.5$) and the spectrum of the
ionizing radiation ($\Gamma=-\alpha_{\rm ox}=2$) given, we follow
the same procedure as in FKCB: we split the wind logarithmically
into a number of radial zones; we employ \verb"XSTAR" to compute the
ionization and opacities/emissivity in each zone along an observer's
LoS.
%
%these are then used to transfer the
%radiation to the next zone along the same direction and repeat the
%procedure over the entire poloidal plane.

\begin{figure}[h]% ------------------------------------- Figure~2
\begin{center}$
\begin{array}{cc}
\includegraphics[trim=0in 0in 0in
0in,keepaspectratio=false,width=3.1in,angle=-0]
{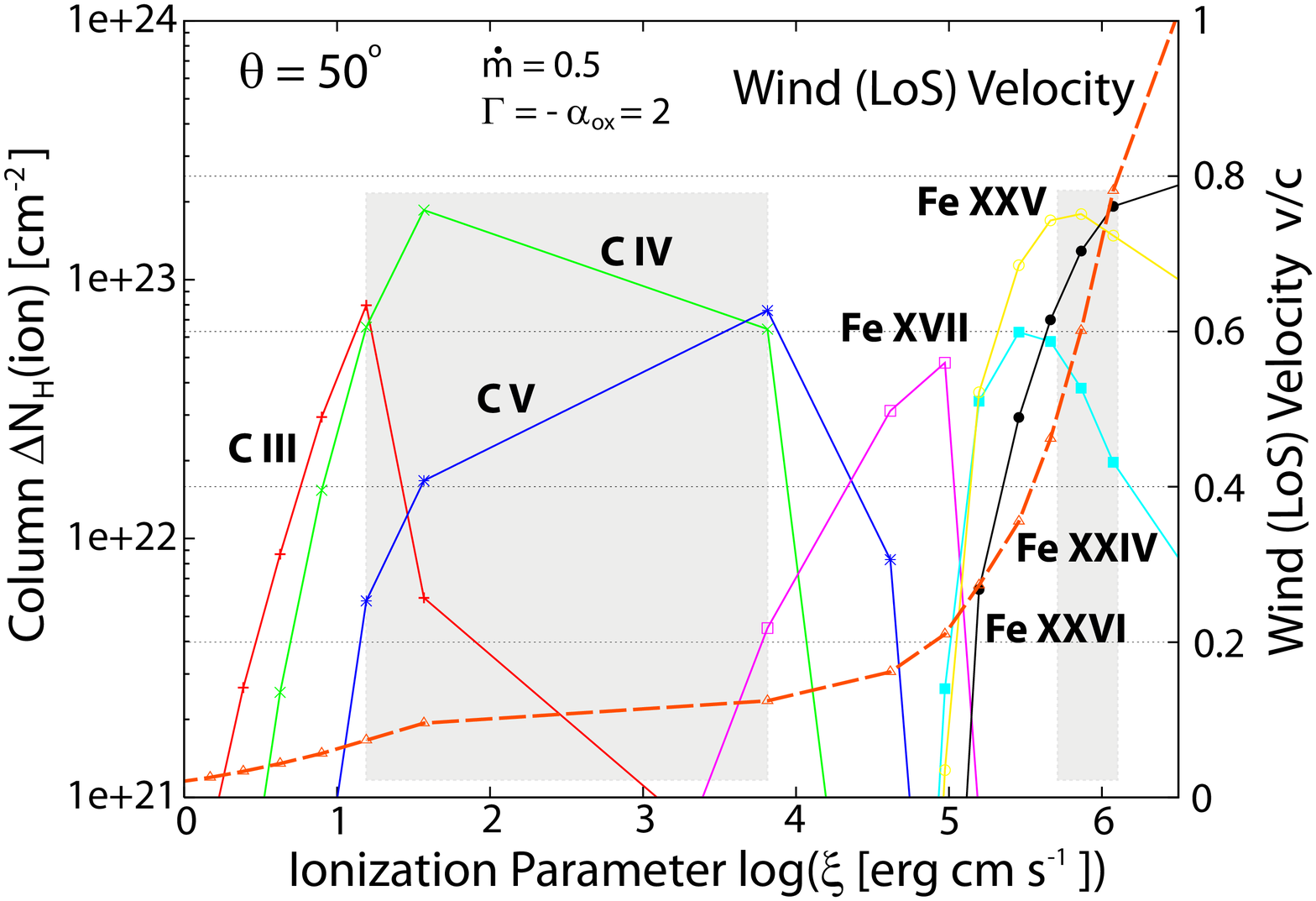} &
\end{array}$
\end{center}
\caption{Simulated distribution of local column densities
$\Delta N_H$ (left ordinate) and the outflow velocity $v$ ({\it dashed} curves; right ordinate)
as a function of ionization parameter $\xi$ for carbon and iron along the $\theta=50\deg$ LoS.
The shaded regions denote the parameter space (in $\xi$ and $v$) for which the local
column is dominated primarily by \civ ~or
\fexxv. } \label{fig:fig2}
\end{figure}

In Figure~\ref{fig:fig2} we show the resulting distribution of the
hydrogen-equivalent column densities $\Delta N_H$ of iron and carbon
for $\theta=50\deg$ as a function of $\xi$ (optimized here to model
the outflows in APM~08279+5255) along with the corresponding LoS
velocity ({\it dashed} curve), to be read on the right vertical
axis. Note the well-defined velocity gradient of the wind with
ionization parameter $\xi$ due to its continuous structure. The
\fexxv\ ions identified through their resonance transitions of
1s-2p/3p obtain their peak hydrogen-equivalent columns of
$N_H$(\fexxv) $\sim 2 \times 10^{23}$ cm$^{-2}$ at ${\rm log} \xi
\gsim 5$ with velocities $v \gsim 0.5c$. On the other hand, the
{\civ} ions (2s-2p transitions) yield $N_H$(\civ) $\sim 10^{23}$
cm$^{-2}$ with corresponding LoS velocities $v \sim 0.1c$,
roughly consistent with the UV/X-ray absorbers in APM~08279+0255
data\footnote[3]{It is conceivable that scatterred/reflected UV
photons could externally fill in the ``true" \civ ~absorption
feature to seemingly reduce its intrinsic column (S. Kraemer,
private communication).} (C09). The obtained ionization structure
directly reflects the spatial positions of these ions; i.e.
$r$(\fexxv)/$r_s$ $\sim 5-40$ and $r$(\civ)/$r_s$ $\sim 200-700$ for
$\theta=50\deg$ (see Fig.1a) assuming a single
LoS\footnote[4]{Different LoS for the UV/Optical and X-ray emitting
regions \citep[e.g.][]{Dai10} could alternatively be modeled with
radiation transfer in a more complex source geometry.}. Note that
the equivalent width (EW) of the modeled \civ ~absorption seems also
to be fairly large due to a wide spread of its peak column
distribution (i.e. $1.5 \lsim \log \xi \lsim 4$) equivalent to
$\Delta v \sim 18,000$ km~s$^{-1}$ in agreement with the typical
width of \civ ~BALs (i.e. $\Delta v \sim 10,000-30,000$
km~s$^{-1}$).

\begin{figure}[h]% ------------------------------------- Figure~3
\begin{center}$
\begin{array}{cc}
\includegraphics[trim=0in 0in 0in
0in,keepaspectratio=false,width=2.9in,angle=-0] {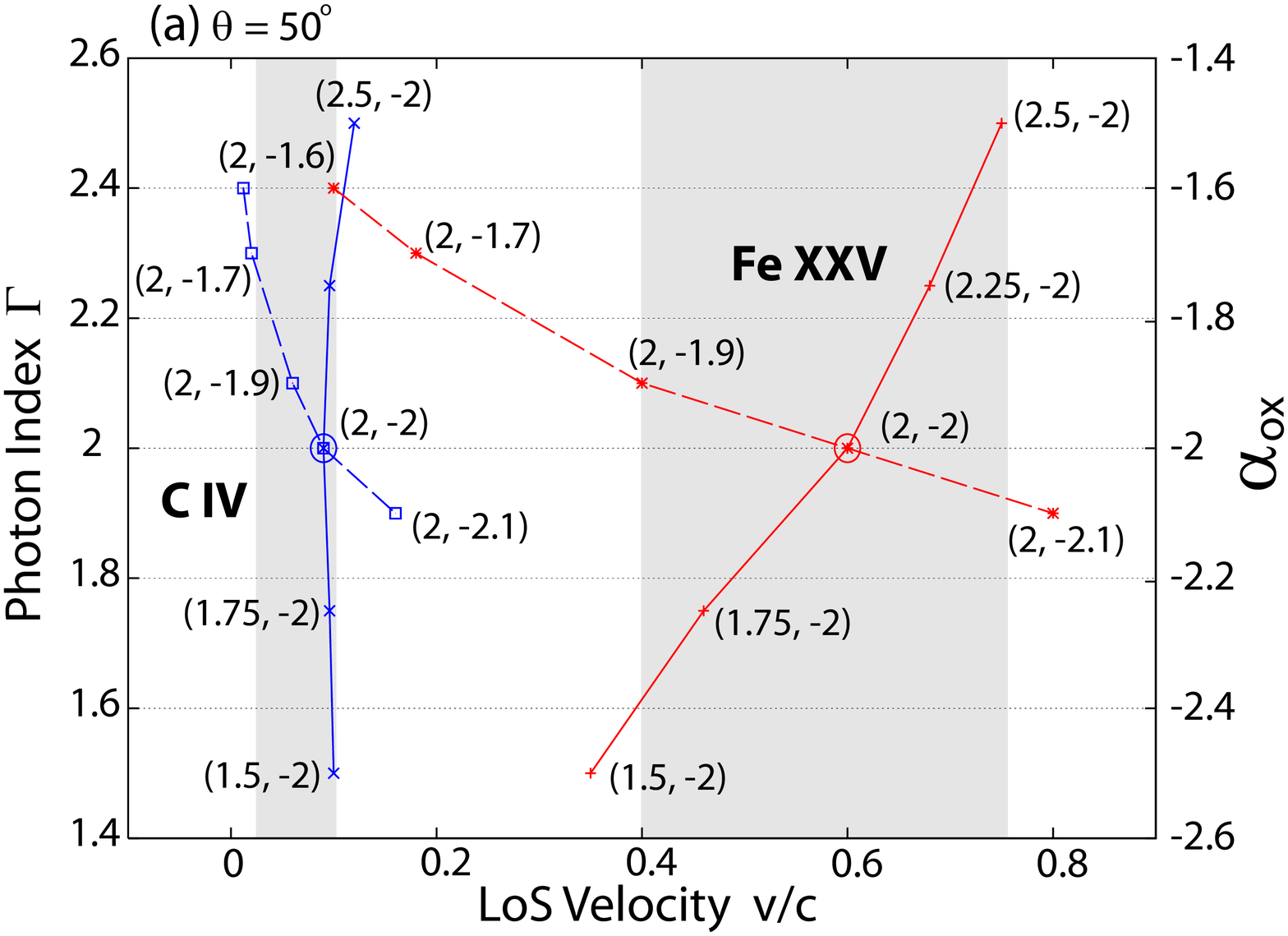} &
\includegraphics[keepaspectratio=false,width=2.9in,angle=-0]
{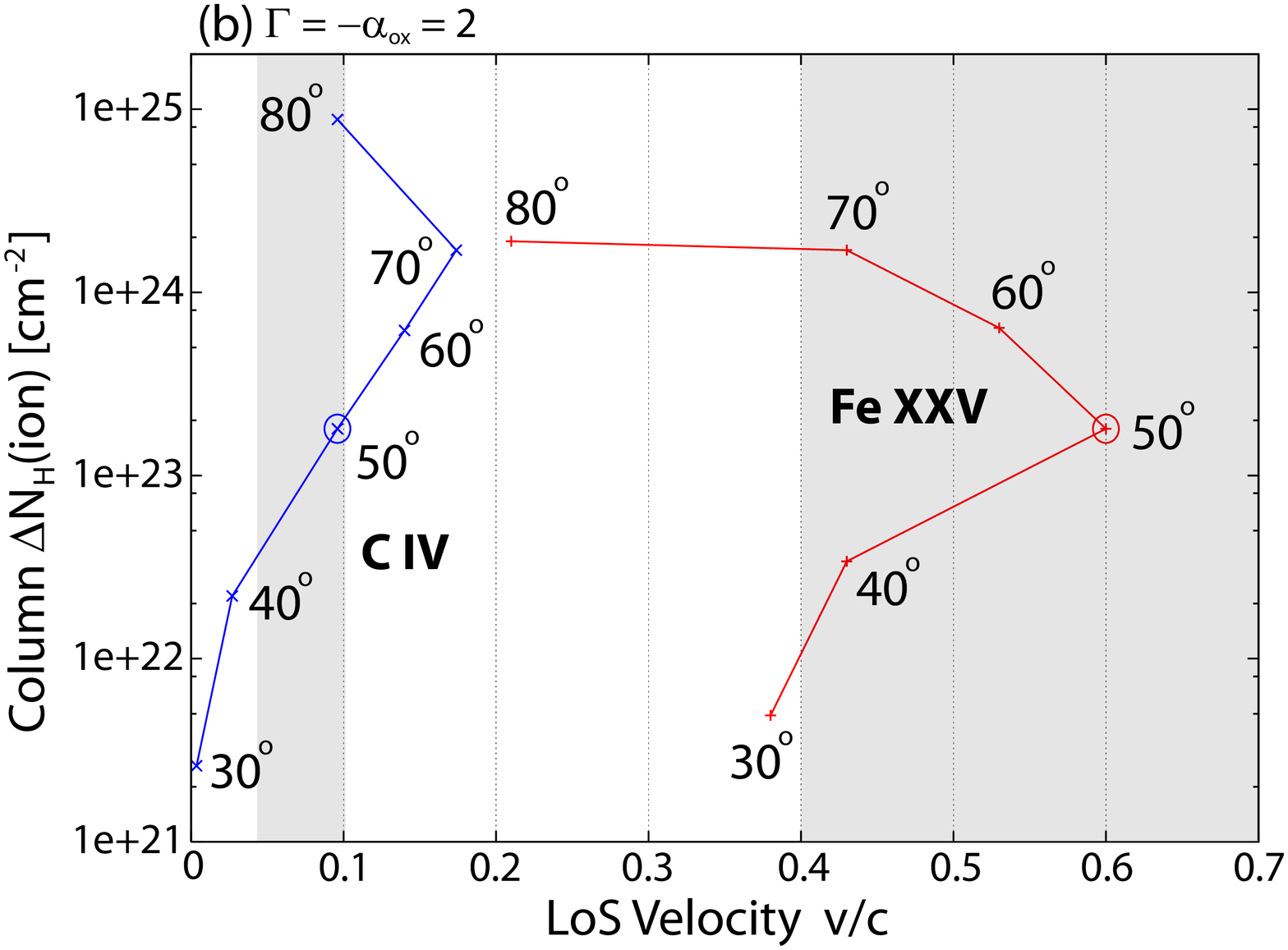}
\end{array}$
\end{center}
\caption{(a) Expected correlation of MHD wind velocity $v$ with $\Gamma$ ({\it solid}
curves; left ordinate) and $\alpha_{\rm ox}$ ({\it dashed} curves; right
ordinate) indicated by ($\Gamma,\alpha_{\rm ox}$) for \civ ~and \fexxv
~with $\dot {m}=0.5$ and $\theta=50\deg$.
(b) Expected correlation of wind velocity $v$ with $\Delta N_H$ and
$\theta$ (indicated by numbers in degree) for \civ ~and \fexxv ~with
$\Gamma=-\alpha_{\rm ox}=2$. Shared regions indicate the observed
velocity dispersions of \civ\ and \fexxv\ for APM~08279+5255. } \label{fig:fig3}
\end{figure}

The difference in the \fexxv ~and \civ ~velocities above from those
in FKCB [$v$(\fexxv) $\sim 3,000$ km~s$^{-1}$ and $v$(\civ) $\sim
300$ km~s$^{-1}$] begs an explanation, considering the scale
invariance of these winds and the similar values of $\theta (\sim
50\deg$) presented in both cases. This can be traced to the
different $\dot m$ and/or ionizing spectrum $F_\nu$. Indeed $\dot m$
here is $\sim 4$ times higher than that in FKCB, leading to a large
increase in $N_H$; furthermore, $\alpha_{\rm ox}$ is smaller here
than in FKCB, with the present value $\alpha_{\rm ox}=-2$ consistent
with the values of UV-luminous QSOs \citep{steffen05}.

These differences affect the \fexxv ~and \civ ~velocities as
follows: The combined increase in $\dot m$ and decrease in
$\alpha_{\rm ox}$ imply  lower ionization of the plasma at the
smallest $r$, despite the fact that the ratio $L/n_o$ does not
change; this is because the luminosity $L$ in the definition of
$\xi$ involves an integral over the spectrum, while the ionization
of the gas is affected mostly by the 2-10 keV X-rays. Therefore, an
increase in $\dot m$ for a given $M$ achieves: (i) an increase in
the plasma column, density and luminosity and (ii) a relative
decrease of the ionizing hard X-ray flux (smaller $\alpha_{\rm
ox}$). With these changes over FKCB, in the present treatment iron
is not fully ionized even at the smallest radii ($r \simeq 10 r_s$),
leading to $v$(\fexxv) $\simeq 0.7c$. This partial ionization of the
plasma, coupled with the increased column, reduces the ionizing flux
that reaches further out into the wind, especially for high
$\theta$, leading to a bootstrap of less ionization and increasingly
higher soft X-ray opacity. Then, because of the ensuing severe
reduction of ionizing photons in the $E \sim 0.1-2$ keV range,
%ionization equilibrium requires that
the \civ\ ions form at smaller radii (higher velocities), so that
the $r^{-2}$ increase of the photon flux with $E \gsim 64$ eV (the
ionization potential of \civ) off-sets the photon depletion due to
photoelectric absorption by the partially ionized plasma.

Recent spectroscopic studies of BAL QSOs have indicated likely
correlations between the maximum outflow velocity of the UV/X-ray
absorbers and the spectral indices (G06;C09). For comparison we show
in Figure~\ref{fig:fig3}a the modeled outflow velocities for \civ
~and \fexxv ~with different values of ($\Gamma, \alpha_{\rm ox}$)
for $\dot m=0.5$ and $\theta=50\deg$. It is seen for $\alpha_{\rm
ox}=-2$ that \fexxv ~velocities correlate strongly with $\Gamma$
({\it solid} curves) allowing for velocities in the range $0.3 \lsim
v({\rm \fexxv})/c \lsim 0.8$ consistent with the X-ray
outflow velocity observations in APM~08279+5255 (C09), while the
\civ ~velocity is virtually unaffected. This is because for the
steeper X-ray spectra fewer ionizing photons are available to
produce highly-ionized species (e.g. \fexxv ~of ionization potential
$\sim 9$ keV) and the relevant ions are found at smaller distances
(higher velocities) than in the case of harder spectra. This does
not affect significantly the overall ionization of the wind leaving
the \civ\ transition at roughly the same distance.
However, a change in $\alpha_{\rm ox}$, affects strongly the maximum
column position of both \fexxv ~and \civ, as described above. For
constant $\Gamma(=2)$, the velocities of both these transitions
correlate strongly with $\alpha_{\rm ox}$ ({\it dashed} curves),
ranging between $0.1 \lsim v$(\fexxv)/c $\lsim 0.8$ and $0.01 \lsim
v$(\civ)/c $\lsim 0.15$ for $-2.1 \lsim \alpha_{\rm ox} \lsim -1.6$
qualitatively consistent with UV data
(\citealt{LaorBrandt02}; G06; \citealt{Fan09}).
Radiation forces are often invoked to explain these correlations
(with X-ray shielding necessary for high velocities, e.g. MCGV;PSK).
In contrast to these models, {\it high-velocity flows are always
present in our model}, but only the steep $\alpha_{\rm ox}(\lsim
-2)$ allows the relevant ions (e.g., \fexxv\ and \civ) to form in
their {\it small-$r$, high-$v$ regions} which are otherwise
overionized (c.f. FKCB).
%
%Recalling
%that no (radio-quiet) Seyfert galaxies (with $-1.6 \lsim \alpha_{\rm
%ox} \lsim -1.4$) appear to exhibit outflow velocities as fast as
%those of bright QSOs [$v({\rm Seyfert})\lsim 3,000$ km~s$^{-1}$;
%see, e.g., \citet{HBK07,HBA10}],
%
%\textbf{While some (narrow-line) Seyferts with $-1.6 \lsim
%\alpha_{\rm ox} \lsim -1.1$ appear to exhibit X-ray outflows with
%low ($v/c \lesssim 0.15$;  \citealt{HBK07,HBA10}) and high ($v/c
%\lesssim 0.3$;\citealt{Tombesi10}) velocities, they are still on the
%lower side of the distribution in comparison with those in BAL QSOs
%with $-2.0 \lsim \alpha_{\rm ox} \lsim -1.6$. Thus, we propose
%$\alpha_{\rm ox}$ as the defining parameter that determines the
%velocities of the UV/X-ray absorption features in AGNs.}
%
While some (narrow-line) Seyferts with $-1.6 \lsim
\alpha_{\rm ox} \lsim -1.1$ appear to exhibit X-ray outflows with
$v/c \lesssim 0.15$ \citep{Dadina05,Tombesi10}, these are
systematically slower [and generally substantially slower
\citep{HBK07,HBA10}] than those of the BAL QSOs with
$\alpha_{\rm ox} \lsim -1.6$. Thus, we propose $\alpha_{\rm ox}$ as
the defining parameter that determines the velocities of the
UV/X-ray absorption features in AGNs.

While the intrinsic MHD wind ionization structure is determined by
$\alpha_{\rm ox}$, the observables, i.e. the velocity widths/shifts
of the ions depend strongly also on the observers' inclination angle
$\theta$.
%
%The ionization-velocity structure of the winds discussed above,
%besides the (dimensionless) accretion rate, $\dot m$, and the
%spectral distribution, $\alpha_{\rm ox}$, involves also the
%observer's LoS angle $\theta$.
%
In Figure~\ref{fig:fig3}b we present the LoS velocity of the \civ
~and \fexxv ~transitions for various $\theta$ with $\Gamma=
-\alpha_{\rm ox}=2$. Because of the specific geometric shape of the
magnetic field lines and ionization equilibria, characteristic ion
velocities vary for different LoS angles [see also Fig.~1a for their
positional transitions  along various LoS from $\theta=80\deg$
(innermost) to $30\deg$ (outermost)].
%
% tangent to the wind velocity vector $\mathbi{v}$.
%
In this fiducial model we find that $v_{\rm max}({\rm \fexxv}) \sim 0.6c$ at $
\theta \sim 50\deg$ and $v_{\rm max}$(\civ) $\sim 0.15c$ at $\theta \sim
70\deg$. At larger angles the velocities decrease but the integrated
columns are so high that it is doubtful these features are
observable. Similar diagrams can be computed for different values of
the parameters ($\dot m, \Gamma, \alpha_{\rm ox}$) and can be directly
compared to observations to assess the fundamental assumptions of
these models.

\section{Summary \& Discussion}

We have demonstrated that purely MHD disk-winds with $n \propto
r^{-1}$, originally proposed to account for the X-ray AMDs in
Seyferts, can also encompass combined {\it high-velocity UV/X-ray
absorber properties} as diverse as those of BAL QSOs, with those of
APM~08279+5255 as a template. This is extremely important in view of
the winds' scale invariance, with the qualitative differences in the
absorber properties between Seyferts and QSOs attributed mainly to
their different $\dot m$ and $\alpha_{\rm ox}$. Given the well
documented correlation of AGN UV-luminosity (a proxy for $\dot m$)
with $\alpha_{\rm ox}$ \citep{LaorBrandt02,steffen05,Fan09}, this
model implies AGN absorber structure that depends essentially on a
{\em single} parameter $\alpha_{\rm ox}$. However, the observables,
e.g., columns and velocities, depend additionally on the LoS angle
$\theta$, reproducing the QSO BALs \civ ~(UV) and \fexxv ~(X-ray)
properties only for sufficiently large $\theta$, as usually
considered.

Our calculations show that \civ\ forms at $r/r_s \simeq 200-700$
with corresponding velocity $v$(\civ) $\lsim 0.1c$. This value,
along with the \fexxv ~velocity $v$(\fexxv) $\simeq 0.6c$ at $r/r_s
\simeq 5-40$, are consistent with those observed in APM~08279+5255
(C09). The shielding of the plasma from the X-rays at $r \gsim 100
r_s$, necessary to produce the high velocity \civ ~absorption in
radiation-driven wind models (MCGV;PSK), is in our case naturally
provided by the {\it faster components of the same wind
%
%which includes also
%high-velocity components
launched from even smaller radii}. Ionization equilibrium as a
result of steeper $\alpha_{\rm ox}$ allows the relevant charge
states to form closer to the central engines where the wind is
faster.

%The correct of the tenets of this model can be
%decided by testing its statistical properties against observation.
% (the absence of low ionization species in this
%object can be accounted by companion which limits the disk radial
%extent the finite radial extent).

For simplicity, we have ignored here a number of physical processes,
e.g. radiation pressure (see PSK) and thermal instability
\citep{KMT,HBK07}, likely to have an impact on our MHD-wind
properties, that need to be implemented \citep[e.g.][]{Proga03}.
However, the broader validity of our models, gauged by the AMD
dependence on $\xi$, will be decided by observations and
quantitative analysis such as those presented in B09. It is also
encouraging that the ionization properties of certain X-ray
absorbers are consistent with this picture and, in fact,
magnetic-driving of disk-winds has been favorably argued for
GRO~J1655-40 \citep[e.g.][]{Miller08} and NGC~4151
\citep[e.g.][]{CK07}, for example.
%
%It is encouraging that the ionization properties
%of the galactic source GRO J1655-40 are consistent with this
%picture, in fact they were used to argue for the magnetic driving of
%its winds \citep{Miller08}.
%
We anticipate the upcoming {\it Astro-H} mission to contribute
significantly to this goal by providing more detail on the Fe-K
component of the wind, and thus to further clarify our picture of
AGN structure.

\acknowledgments

Authors are grateful to the anonymous referee for inspirational suggestions. K.F.
and D.K. would like to thank T. Kallman for insightful discussions, and G. Chartas, S. Kraemer, F. Tombesi and J. Turner for their constructive comments.

\end{document}